# Influence of Metal-Graphene Contact on the Operation and Scalability of Graphene Field-Effect-Transistors

Pei Zhao[1], Qin Zhang[1,2], Debdeep Jena[1*], and, Siyuranga O. Koswatta[3]

[1] Department of Electrical Engineering, University of Notre Dame, Notre Dame, IN 46556, USA
[2] CMOS and Novel Devices Group, Semiconductor Electronics Division, National Institute of Standards and Technology (NIST), Gaithersburg, MD 20899, USA
[3] IBM Research Division, T. J. Watson Research Center, Yorktown Heights, NY 10598, USA

## Abstract

We explore the effects of metal contacts on the operation and scalability of 2D Graphene Field-Effect-Transistors (GFETs) using detailed numerical device simulations based on the non-equilibrium Green's function formalism self-consistently solved with the Poisson equation at the ballistic limit. Our treatment of metal-graphene (M-G) contacts captures: (1) the doping effect due to the shift of the Fermi level in graphene contacts, (2) the density-of-states (DOS) broadening effect inside graphene contacts due to Metal-Induced-States (MIS). Our results confirm the asymmetric transfer characteristics in GFETs due to the doping effect by metal contacts. Furthermore, at higher M-G coupling strengths the contact DOS broadening effect increases the on-current, while the impact on the minimum current ($I_{min}$) in the off-state depends on the source to drain bias voltage and the work-function difference between graphene and the contact metal. Interestingly, with scaling of the channel length, the MIS inside the channel has a weak influence on $I_{min}$ even at large M-G coupling strengths, while direct source-to-drain (S → D) tunneling has a stronger influence. Therefore, channel length scalability of GFETs with sufficient gate control will be mainly limited by direct S → D tunneling, and not by the MIS.

[*] djena@nd.edu

## I. INTRODUCTION

Graphene is a two dimensional, zero bandgap, material with carbon atoms arranged in a honeycomb lattice [1],[2]. A finite bandgap can be obtained through quantum confinement by cutting 2D graphene into strips as graphene nanoribbons (GNRs). High quality GNRs with acceptable edge uniformity, however, pose a technological challenge. On the other hand, monolayer 2D graphene can be achieved by means of mechanical exfoliation of graphite, high temperature sublimation of silicon from SiC substrates [3], or CVD growth on metal substrates [4],[5]. Although 2D monolayer graphene lacks a band gap, it still shows promising potential for applications in high frequency devices that do not require a high on/off ratio as demanded by digital logic [6]-[10]. Recently, many theoretical and experimental works have studied the microscopic physics of metal-graphene contacts, which show that metal contacts could play a critical role on the device performance [11]-[22]. In this regard, theoretical studies based on the density functional theory (DFT) show that when graphene is covered by a metal electrode, the Fermi level of graphene underneath will be shifted [11],[12]. This behavior has also been observed in experiments [13],[14]. On the other hand, the detailed influence of M-G contacts on the transport properties, and more importantly, on the scalability of GFETs has not been addressed yet.

In this work, we study the effect of M-G coupling on the operation and the scalability of 2D GFETs. Even though previous simulation studies have explained the asymmetric transport characteristics in electrostatically doped graphene p-n junctions [23], experiments have indicated that M-G contacts themselves may also lead to asymmetric conduction in 2D GFETs [13],[14]. In this work, we confirm the latter observation by using a self-consistent 2D electrostatic solution of the GFET geometry [24] which captures the metal-induced doping effect. M-G coupling could also lead to metal induced states inside the graphene channel, which is similar in origin to the metal-induced-gap-states in conventional metal-semiconductor contacts [25]. Experiments have also confirmed that the impact of metal contacts on the channel potential extends into the channel for several hundred nanometers [20]-[22]. Here we consider the influence of MIS on the scalability of GFETs and provide detailed insights into the impact of metal contacts on GFET characteristics.

## II. DEVICE MODEL AND THE SIMULATION APPROACH

The modeled device is shown in Figure 1. The channel is assumed to be uniform graphene with width $W$ of 150nm. At $W = 150$nm, the current density (in mA/μm) is similar to the 2D analytical result, which justifies the effective 2D limit of the modeled GFET. The top gate insulator is $t_{ox} = 1.5$nm thick, and has a dielectric constant $\varepsilon_{ox} = 20$. With such excellent dielectric assumptions short channel effects can be avoided in our model, and we can focus on the effects from the M-G contacts (the impact of the oxide thickness will be discussed below). The SiO$_2$ substrate is 50nm in thickness, and connected to the ground. The simulated area is only the channel region (dashed rectangle in Figure 1) with fixed boundary conditions at the source and the drain. The electrostatic solution procedure is described in [24]. The effective oxide thickness (EOT) of our modeled device is only 0.3nm, thus the quantum capacitance ($C_Q$) corresponding to the graphene DOS cannot be ignored. Our simulations self-constantly solve both the transport and the electrostatic parts, thus the effect of quantum capacitance is naturally captured. In other words, the total gate capacitance, $C_G$, is not simply equal to $C_{ox}$, but more generally to, $C_G = (C_{ox} C_Q)/(C_{ox} + C_Q)$.

Here, the tight-binding model for ballistic transport in the channel is assumed, and it is solved using a mode-space based non-equilibrium Green's function formalism [26]. The mode-space approach significantly reduces the computational cost while still maintaining the accuracy, as has already been demonstrated in simulations of MOSFET [27], carbon nanotube FET [28], and GNR FET [29]. In our calculations, we assume that the potential variation along the channel width direction is negligible. Based on this approximation, the electrostatics is solved in 2D, and the mode-space method yields accurate results.

In this work, the contact regions are assumed to be semi-infinitely long, and comprise of a metal layer deposited on top of the graphene layer [11],[16]. In this M-G hybrid system (dotted rectangle in Figure 1), the Fermi level of the graphene underneath is shifted and the DOS is broadened due to the M-G coupling [11],[12]. The Fermi level shift in the contact regions is modeled by, $\Delta E_{contact} = E_F - E_{Dirac}$, where $E_F$ ($E_F = 0\text{eV}$) is the Fermi level and $E_{Dirac}$ is the Dirac point inside the graphene contact regions. The broadening of DOS is captured by a phenomenological approach [15]-[18] with a M-G coupling strength of $\Delta$ (in meV) that can reproduce the *ab initio* simulation results [17].

The effect of M-G contacts on the channel region is captured through a contact self-energy function, $\Sigma_{S,D} = \tau g_S \tau^{\dagger}$, where $\tau$ is the coupling matrix between the contact and the channel ($\tau^{\dagger}$ is its Hermitian conjugate), and $g_S$ is the surface Green's function of the contact, $g_S(E) = [(E+i\Delta)I - H_{contact}]^{-1}$ [16]. Here, $H_{contact}$ is the contact Hamiltonian matrix, and $I$ is the identity matrix. In the mode-space approach, $g_S$ of the $q$th mode can be analytically expressed as:

$$g_{sq} = \frac{(E-U_1+i\Delta)^2 + b_{1q}^2 - b_{2q}^2 + \sqrt{[(E-U_1+i\Delta)^2 + b_{1q}^2 - b_{2q}^2]^2 - 4(E-U_1+i\Delta)^2 b_{1q}^2}}{2b_{1q}^2(E-U_1+i\Delta)}.$$

The source contact self-energy function for the $q$th mode is $\Sigma_{Sq}^{1,1} = (b_{1q})^2 g_{sq}$, where $b_{1q} = t_0$ and $b_{2q} = 2t_0 \cos(q\pi/(n+1))$ are the coupling parameters in the $q$th mode 1D sublattice, $t_0$ is the nearest neighbor tight binding parameter, $n$ is the number of carbon atoms in the width direction, and $U_1$ is the electrostatic potential at the source end [29]. A similar expression applies for the drain contact self-energy function, $\Sigma_{Dq}$, with $U_1$ being replaced by the potential at the drain end. The retarded Green's function for $q$th mode is then determined by $G_q(E) = [(E+i0^+)I - H_q - \Sigma_{Sq} - \Sigma_{Dq}]^{-1}$, where $H_q$ is the channel Hamiltonian matrix for the $q$th mode

$$[H_q] = \begin{bmatrix} U_1 & b_{2q} & & & \\ b_{2q} & U_2 & b_{1q} & & \\ & b_{1q} & U_3 & b_{2q} & \\ & & \ldots & \ldots & \end{bmatrix}.$$

Here $U_i$ is the electrostatic potential at the $i$th atom in the $q$th mode. The source (drain) local-density-of-states (LDOS) within the channel are computed as $LDOS_{S(D)} = G_q \Gamma_{S(D)} G_q^{\dagger}/2\pi$, where $\Gamma_{S(D)} = i(\Sigma_{S(D)} - \Sigma_{S(D)}^{+})$ is the energy broadening due to the source (drain) contact. The total LDOS within the channel is given by the summation of $LDOS_S$ and $LDOS_D$. Finally, the channel current is computed from $I_{DS}=(2e/h)\int T(E)(f_S(E)-f_D(E))dE$, where $f_{S/D}$ are the source/drain Fermi-Dirac distribution functions, respectively, and $T(E)=\sum_q T_q(E)$ is the total transmission coefficient with $T_q$ being the transmission of the $q$th mode.

### III. EFFECT OF METAL-GRAPHENE CONTACT ON THE CHANNEL

To explore the influence of M-G contact on the channel, we plot the energy-position-resolved channel LDOS in logarithmic scale in Figure 2 for (a) $\Delta = 0$meV and (b) $\Delta = 50$meV. The device structure is same as shown in Figure 1 with $V_{DS}= 0$V and $V_{GS}= 0.1$V. The potential profiles of the channel Dirac point (dashed lines) are the self-consistent results. The channel Dirac point at source and drain ends equal to 0.2eV, which shows the p-type doping effect inside the contact with $\Delta E_{contact} = -0.2$eV. When $\Delta = 0$meV, LDOS near the contact Dirac point energy level (0.2eV) is very low (darker color). As mentioned in Sec. II, states within the channel are given by $LDOS_{S(D)} = G_q \Gamma_{S(D)} G_q^\dagger/2\pi$. When $\Delta = 0$meV, the contact broadening effect $\Gamma_{S(D)} \approx 0$ leads to a negligible broadening of the channel states. Even though states do exist within the channel, those states with infinitesimal broadening are not visible in the energy-position-resolved LDOS. However, when $\Delta = 50$meV, $\Gamma_{S(D)}$ is large which gives obvious broadening effect of the channel states. Thus energy-position-resolved LDOS clearly shows states near the energy level of the contact Dirac point in Figure. 2 (b). Near the gate controlled channel Dirac point at about -0.09eV, *contact induced evanescent states* can be observed penetrating into the channel. Those states originate from the wave-functions incident from the contacts. Because graphene has a zero bandgap, those evanescent states can penetrate for a long distance.

Figure 2 (c) and (d) plot LDOS *vs.* energy at the source end of the channel, and at the middle of the channel for two Δ values. It is clear that at Δ = 0meV, a negligible $\Gamma_{S(D)}$ will lead to unobservable states at the energy level of the contact Dirac point. M-G coupling can broaden the zero-DOS near the contact Dirac point, but the effect of broadening is weak if the intrinsic graphene DOS is large (sketch of intrinsic DOS solid line and broadened DOS dashed line in Figure 2 (b)). Along the channel Dirac point around -0.09eV, MIS in channel is negligible, thus LDOS does not have a dependence on Δ. Contact induced evanescent states will affect the channel LDOS, but the effect of MIS due to M-G coupling Δ is trivial. Golizadeh-Mojarad's work [15] shows that without any bias, contact Dirac point is at the same energy level as the channel Dirac point. MIS along the channel Dirac point show strong dependence on M-G coupling Δ. This conclusion is valid if no bias is added and contact Fermi level is fixed at Dirac point. Our work provides a complete description since the metal induced contact doping effect is captured, and the potential profile is solved self-consistently at any given bias. The calculated LDOS inside the channel is the summation of the intrinsic graphene DOS and states due to contact incident wave-function penetration. Inset figure of Figure 2 (c) shows that, at $L_{ch}$ = 15nm, contact incident wave-functions increase the channel LDOS, which has a large influence on the minimum current as discussed later.

## IV. RESULTS AND DISCUSSION

Figure 3 shows the calculated transfer characteristics (T = 300K) at low drain bias $V_{DS}$ = 0.1V with different Fermi level alignments $\Delta E_{contact}$ in the contact regions. The M-G coupling strength is taken to be Δ = 50meV. The effects of different Δ's are discussed later. If $\Delta E_{contact}$ = 0eV, graphene contact is not doped by metal, the transfer curve is symmetric as shown in Figure 3 (a). The minimum conduction point is located at gate voltage $V_G = V_{DS}/2$. In (b) $\Delta E_{contact} = -0.2$eV (p-type doping), a typical value for Au contacts [11]. At negative $V_G$, carriers can directly go through the channel. At positive $V_G$, the electrons need to go through the channel Dirac point (see schematic potential of the Dirac point). Because DOS near the Dirac point is very low, it suppresses the carrier injection from contact to channel. Thus, the positive current branch is reduced compared to the negative branch. The complete transfer curve shows a clear asymmetric behavior. We point out that the gate voltage at which the minimum conduction point occurs is also slightly shifted due to the asymmetric barriers at the contacts. When $\Delta E_{contact}$ is positive as in Figure 3 (c), an effective n-type doping is introduced by the metal contacts. A similar asymmetric behavior with the positive current branch being greater than the negative branch is seen in that case. Another interesting feature is, compared with $I_{DS}$ - $V_{GS}$ in Figure 3 (a), contact doping effect increases the on current as shown in Figure 3 (b) and (c).

The effect of different coupling strengths $\Delta$ at $V_{DS} = 0.1$V is shown in Figure 4. The rigorous explanation of the values used for $\Delta$ is related to the M-G hybrid system, which is beyond the focus of this work (see [11],[12] for details). For a comparison, we define three $\Delta$ values here; $\Delta = 0$eV for the intrinsic graphene, $\Delta = 8$meV for weak M-G coupling, and $\Delta = 50$meV for strong coupling. Figure 4 (a) and Figure 4 (b) show the transfer characteristics at channel length $L_{ch} = 300$nm and 15nm, respectively. We observe that the on-current *increases* for larger $\Delta$, which can be understood by looking at the impact of $\Delta$ on $T(E)$ in the on-state (Figure 4(c) center panel). First of all, it is seen that at $\Delta = 0$eV there are three distinct minimum points in $T(E)$ corresponding to the zero-DOS Dirac point position inside the channel, source, and drain regions, respectively. On the other hand, at larger $\Delta$, $T(E)$ near the source and the drain Dirac points increases due to metal induced DOS broadening inside the M-G contact regions, which ultimately enhances the current transport (Figure 4(c right panel)). Interestingly, the minimum current $I_{min}$, however, does not show a dependence on $\Delta$. Figure 4 (d) shows the potential profiles of the Dirac point (left), $T(E)$ (center) and the energy-resolved current density $J(E)$ (right) at the minimum conduction point of $L_{ch}$ = 15nm device. Although the DOS of graphene contact is broadened by the metal contact, $T(E)$ in the *current carrying* energy window between the source Fermi level $E_{FS}$ and the drain Fermi level $E_{FD}$ remains the same for different $\Delta$ values. At $L_{ch}$ = 15nm, *direct* S → D tunneling is the dominant factor controlling $I_{min}$. Furthermore, it is necessary to point out that the above interesting features due to various $\Delta$ still remain valid with other oxide thickness and dielectric values, since the aforementioned effects are mainly due to contact DOS broadening. Here, another transfer characteristics with $t_{ox}$ = 5nm, $\varepsilon_{ox}$ = 9 and $\Delta$ = 50meV (solid lines) are shown in order to compare different EOT values. When EOT increases, the current decreases. The current reduction is due to the loss of $C_G$ and longer effective electrostatic scaling length at larger EOT [24]. At large positive and negative $V_{GS}$, $C_Q$ is large, since $C_G = (C_{ox} C_Q)/(C_{ox} + C_Q)$, $C_G$ is dominant by $C_{ox}$. When EOT increases, $C_{ox}$ decreases, thus the loss of $C_G$ is the main reason leads to current reduction. On the other hand, at small $V_{GS}$, $C_Q$ is small and dominant. Drop of $C_G$ is small and longer effective electrostatic scaling length further limits the current. In addition, when EOT increases, the longer effective electrostatic scaling length further increases the asymmetry $I_{DS}$ - $V_{GS}$ characteristics. Short channel effects can also be observed with large EOT, at $L_{ch}$ = 15nm, the minimum conduction point shifts about 0.15V while at $L_{ch}$ = 300nm such shifts are very small.

Figure 5 (a) and (b) show a comparison of the effect of different coupling strengths Δ at large $V_{DS}$ = 0.3V. When Δ = 0eV, in addition to the primary minimum conduction point, a distortion appears in the $I_D$-$V_{GS}$ characteristics. The two minima are due to the misalignment of Dirac point of channel and drain region. With $V_{DS}$ = 0.3V, Drain Dirac point locates between $E_{FS}$ and $E_{FD}$, which corresponding to primary minimum conduction point. Channel Dirac point is controlled by gate. At $V_{G1}$=-0.05V all carriers pass through the channel hole cone, and at $V_{G2}$=0.1V carriers move through the channel electron cone. But at $V_G$=0.025V, both electron and hole cones are involved, thus channel Dirac point leads to a local minimum. When Δ = 50meV, DOS near contact Dirac point are broadened. The only minimum conduction point is due to channel Dirac point and the distortion disappears. A recent experimental work reports the presence of this type of distortion before annealing, and the disappearance of the distortion after annealing [30]. Our model explains this behavior without resorting to new postulates (such as charge depinning at metal contacts as proposed in [30] ). Before annealing, the M-G interface is not clean and the coupling is weak; after annealing better M-G coupling is achieved and the distortion disappears. In contrast to the low $V_{DS}$ case in Figure 4, $I_{min}$ at large $V_{DS}$ shows a dependence on Δ. Figure 5 (d) shows the internal transport properties near the minimum conduction point of $L_{ch}$ = 15nm GFET. In this case, at large $V_{DS}$ the Dirac point in the drain region is located between $E_{FS}$ and $E_{FD}$. The broadened DOS in the drain contact at larger Δ increases $J(E)$, and thus the $I_{min}$. On the other hand, $T(E)$ near the channel Dirac point energy at about -0.3eV, *does not* show a dependence on Δ. Thus, the increase in $I_{min}$ can be attributed to the DOS broadening in the contacts. Here, we point out that the "large $V_{DS}$" condition is determined by $|V_{DS}| > |(E_F-E_D)/q|$ which would increase $I_{min}$ at larger $\Delta$ values.

The performance of GFETs upon scaling of the channel length is crucial for technology scaling, which is discussed in Figure 6. Our model assumes a high-k insulator to obtain superior electrostatic gate control to avoid the short channel effect. In Figure 6, $\Delta$ = 50meV, $I_{on}$ remains almost the same for different channel lengths, but $I_{min}$ increases at short $L_{ch}$. At short channel lengths, $I_{min}$ is mainly affected by direct S → D tunneling. Here, when $L_{ch}$ is reduced from 150nm to 40nm, because the probability for S → D tunneling rises, $I_{min}$ increases about 10%. When $L_{ch}$ = 15nm, S → D tunneling becomes more severe, $I_{min}$ increases 1.5 times compared with $L_{ch}$ = 150nm. A similar scaling behavior persists even at $V_{DS}$ = 0.3V, $I_{min}$ increases 20% when the channel length shrinks from 150nm to 15nm (not shown).

The $I_D$ vs. $V_{DS}$ characteristics is shown in Figure 7 for (a) $V_{GS}$ > 0V and (b) $V_{GS}$ ≤ 0V. $\Delta E_{contact}$ = 0.5eV is assumed as a degenerate n-type contact. $I_{DS}$-$V_{DS}$ characteristics at $V_{GS}$ = 0.2V and $V_{GS}$ = 0.4V in Figure 7 (a) shows a kink characteristics due to an ambipolar channel, which is also observed in the experiments [6]. In the unipolar regime, where $V_{DS}$ < $V_{DS,kink}$, the GFET shows saturating characteristics [6]. With $V_{GS}$ > 0V, the source/channel/drain are all n-type and comprise an n-n-n type structure, electrons will directly transport through the channel, and thus the dependence of the contact DOS and M-G coupling strength $\Delta$ is small. With $V_{GS}$ ≤ 0V in (b), at weak coupling as $\Delta$ = 0meV, a saturation behavior has been observed. The reason for this saturation behavior is because $V_{GS}$ ≤ 0V leads to p-type channel and n-p-n type structure, and the n-type contact needs to inject electron into the channel. When $\Delta$ = 0meV, low contact DOS near the Dirac point limits the carrier injection from the contacts. With large $\Delta$, contact DOS is broadened and the saturation behavior disappears, which is similar to large $\Delta$ increasing the on current in Figure 5.

Finally, we want to discuss the device performance optimization considering the influence of M-G contact. In Figure 5, we point out that the broadening of DOS in contacts will increase the $I_{min}$ at large $V_{DS}$. Thus, to avoid $I_{min}$ current increase due to DOS broadening, drain bias needs to satisfy $|V_{DS}| < |(E_F-E_D)/q|$. On-current is benefited from strong M-G coupling as shown in Figure 4. We have also explored the dependence on different $\Delta E_{contact}$ (not shown). If $|\Delta E_{contact}|$ is large, metal induced doping effect becomes stronger, the on-current will increase and $I_{min}$ can be controlled with appropriate $V_{DS}$. Metals with large work function difference compared to graphene could be a good candidate to provide large $|\Delta E_{contact}|$ values [11],[12]. The other important issue is that the gate electrostatics play a crucial role. With only the back gate, very long band bending lengths (long effective electrostatic scaling length) near the contacts has been experimentally observed [20]-[22]. Our simulation here is based on excellent top gate electrostatics, thus the band bending length is small, which helps to control the $I_{min}$ while increasing the on-current.

A recent paper has explored the performance of 2D GFETs under ballistic limits [31]. In Ref. [31], the contact self-energy is assumed to be constant and independent of energy, which is a good approximation when metal destroys the linear dispersion of graphene [11]. Our model uses a different approach to describe the contact, where the contact self-energy depends on the coupling strength $\Delta$. Using $\Delta$ as a variable parameter, additional interesting effects have been discussed in this paper. When Δ is small, the linear dispersion of graphene still exists [11,17]. If the coupling strength Δ further increases to about 0.3eV, the graphene contact will become metallic-like, and we can also obtain similar results as the pure metal contact case [18].

## V. CONCLUSION

In this work, we used detailed numerical simulations to investigate the impact of metal contacts on the operation of GFETs with electrostatically well-designed top gates. The metal contacts introduce two effects: (1) the Fermi level of graphene underneath the metal is shifted resulting in asymmetric transfer characteristics; (2) the DOS of graphene inside the contacts is broadened. Δ is introduced to describe the broadening of the DOS inside the M-G contacts. Based on our results, a weak coupling of metal contacts can cause a distortion of the transfer characteristics, which disappears at strong coupling strengths. Large Δ broadens the contact DOS and increases the $I_{on}$ but does not affect $I_{min}$ at low $V_{DS}$. At large $V_{DS}$, i.e. $|V_{DS}| > |(E_F-E_D)/q|$, DOS broadening (MIS inside the contacts) increases both $I_{on}$ and $I_{min}$. With scaling of channel length, direct S → D tunneling is the crucial factor that increases $I_{min}$ at short channel lengths.

**Acknowledgement.** The author would like to thank Dr. Zhihong Chen of IBM T. J. Watson Research Center, for many fruitful discussions.

FIGURE CAPTIONS

Figure 1: Structure of modeled device. The graphene channel is fully covered by the top gate. $t_{ox}$=1.5nm and $\varepsilon_{ox}$=20, the SiO$_2$ substrate is 50nm thick and is connected to ground. This dielectric assumption can avoid short channel effects, and it is used as the nominal condition.

Figure 2: Energy-position-resolved channel LDOS at $V_{DS}$=0V and $V_G$=0.1V, $\Delta E_{contact}$=-0.2eV. In (a) $\Delta$ = 0meV, LDOS along contact Dirac point at 0.2eV are unobservable due to contact induced a negligible broadening $\Gamma_{S(D)} \approx 0$. (b) When $\Delta$ = 50meV, broadening effect $\Gamma_{S(D)}$ is large, then states near 0.2eV is broaden and can be observed. (c) shows the LDOS vs. energy plot at source end of channel and (d) at middle of channel for different $\Delta$. Inset figure show the channel LDOS at -0.09eV, comparing with $L_{ch}$ = 300, when $L_{ch}$ = 15nm LDOS along channel increase due to contact induced states.

Figure 3: Transfer characteristics under different $\Delta E_{contact}$. (a) without any shift of Fermi level in the contact region, symmetric transfer characteristics can be seen. When the Fermi level of contact regions of graphene is shifted due to the metal contact as shown in (b) $\Delta E_{contact}$=-0.2eV and (c) $\Delta E_{contact}$=0.2eV, asymmetric transfer characteristics are observed.

Figure 4: Effect of different coupling strength $\Delta$ at $V_{DS}$=0.1V. On current increases at larger $\Delta$ for both (a) $L_{ch}$=300nm and (b) $L_{ch}$=15nm, the solid lines are the transfer characteristics with $t_{ox}$=5nm, $\varepsilon_{ox}$=9 and $\Delta$ = 50meV. When EOT increase, minimum conduction point shifts due to short channel effect. (c) The increase of on current corresponds to the broadened DOS of the graphene contact. However $I_{min}$ doesn't change with $\Delta$ even at $L_{ch}$=15nm. (d) $T(E)$ between $E_{FS}$ and $E_{FD}$ do not depend on $\Delta$, suggesting direct S → D tunneling is the dominant factor.

Figure 5: Effect of different coupling strength $\Delta$ at $V_{DS}$=0.3V. Besides the minimum conduction point at about 0.2V, a distortion appears when $\Delta$ = 8meV (triangle) and $\Delta$ = 0meV (pentagram). Considering intrinsic graphene, (c) shows the source, drain and the channel DOS (cartoon) in series decide how total carriers transport through the channel. When electron and holes cones are both involved in the transport as at $V_G$=0.025V, a local minimum is given. As $\Delta$ increases, DOS of graphene contact are broadened and distortion disappears. (d) For $L_{ch}$=15nm at the minimum conduction point $V_G$=0.3V, the drain Dirac point is located between $E_{FS}$ and $E_{FD}$, large $\Delta$ broaden the DOS at the drain contact, increasing the $T(E)$ and $I_{min}$.

Figure 6: Effect of channel length scaling at $V_{DS}$=0.1V. A high-k insulator is used to avoid the short channel effect. $I_{on}$ keeps the same when $L_{ch}$ is scaled down.

As *L*$_{ch}$ is reduced from 150nm to 40nm, *I*$_{min}$ increase 10% due to the direct S/D tunneling. When *L*$_{ch}$=15nm, direct S → D tunneling becomes more severe and *I*$_{min}$ increase by about 1.5 times.

Figure 7: Effect of coupling strength Δ on the *I*$_{DS}$ vs. *V*$_{DS}$ characteristics. Δ*E*$_{contact}$ = 0.5eV as the heavily doped n-type contact, and *L*$_{ch}$ = 100nm. The *I*$_D$ vs. *V*$_D$ is grouped into two areas, (a) *V*$_{GS}$ > 0V and (b) *V*$_{GS}$ ≤ 0V. (a) shows a kink characteristics with a saturation region due to an ambipolar channel. Influence of Δ is small. With V$_G$ ≤ 0V in (b), with weak coupling as Δ=0meV, a saturation behavior has been captured. With large Δ, saturation behavior disappears.

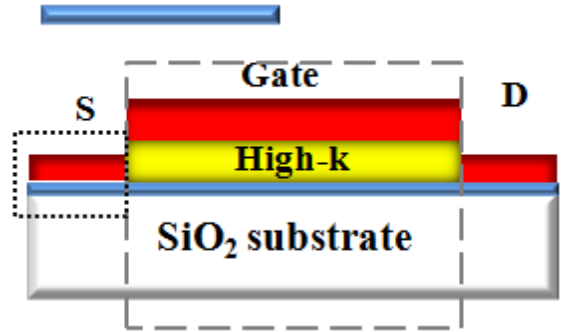

Figure 1

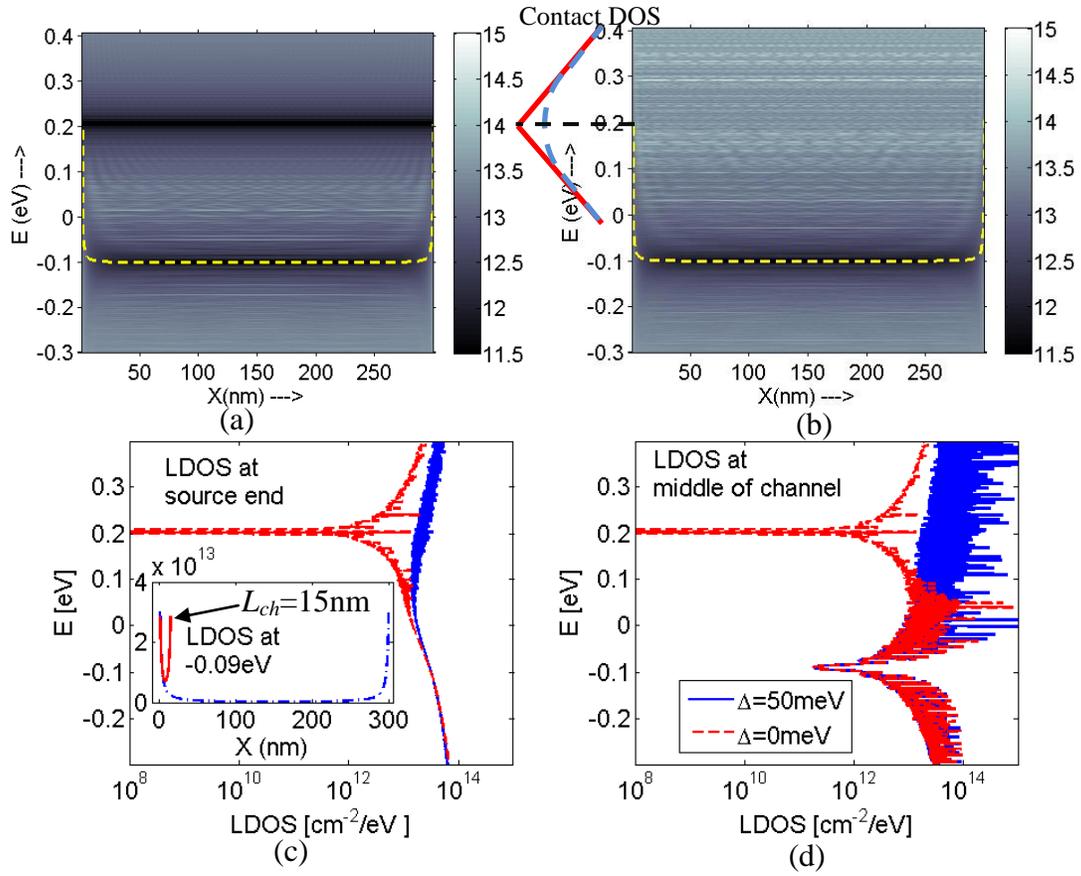

Figure 2

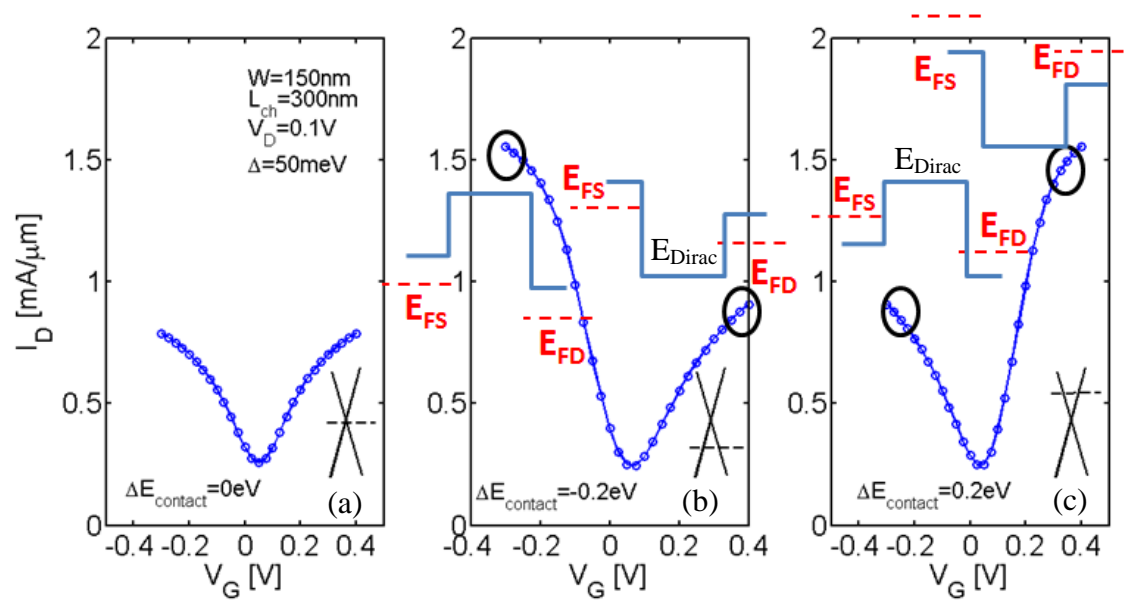

Figure 3

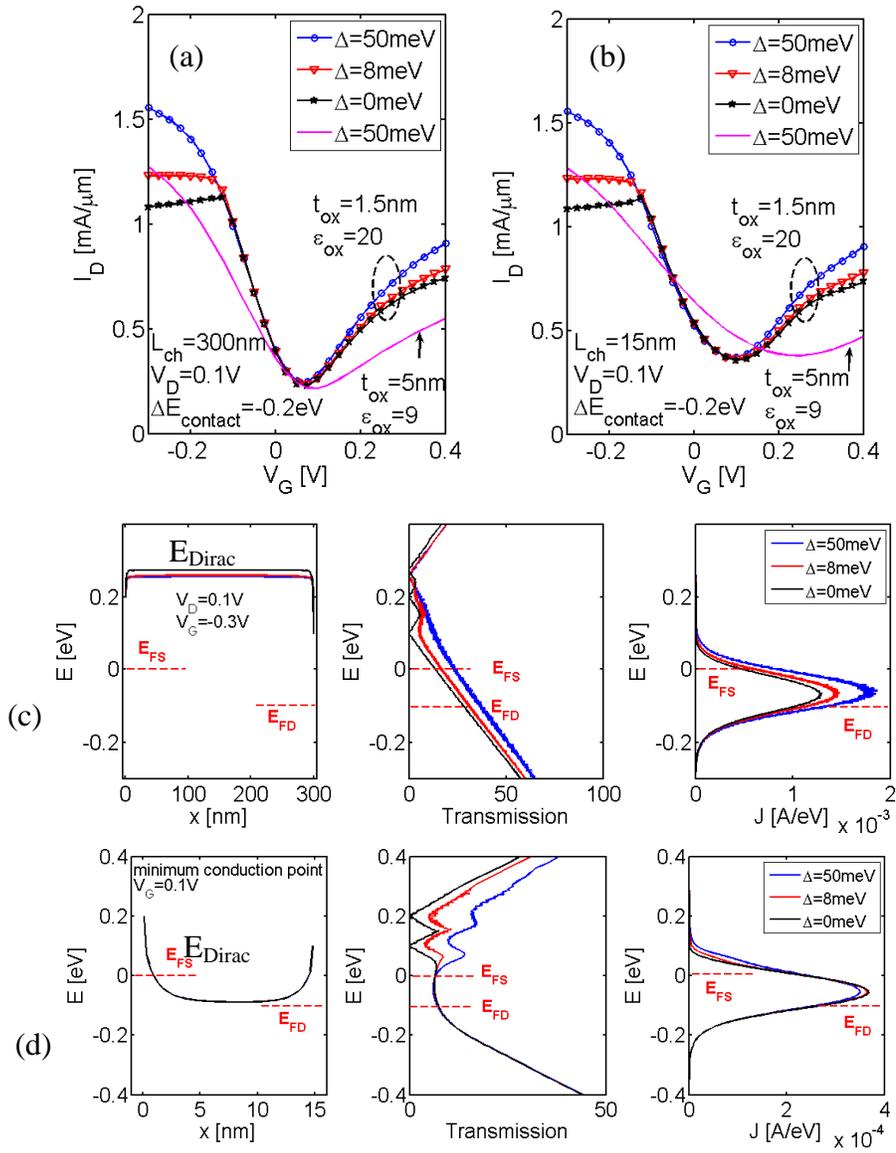

Figure 4

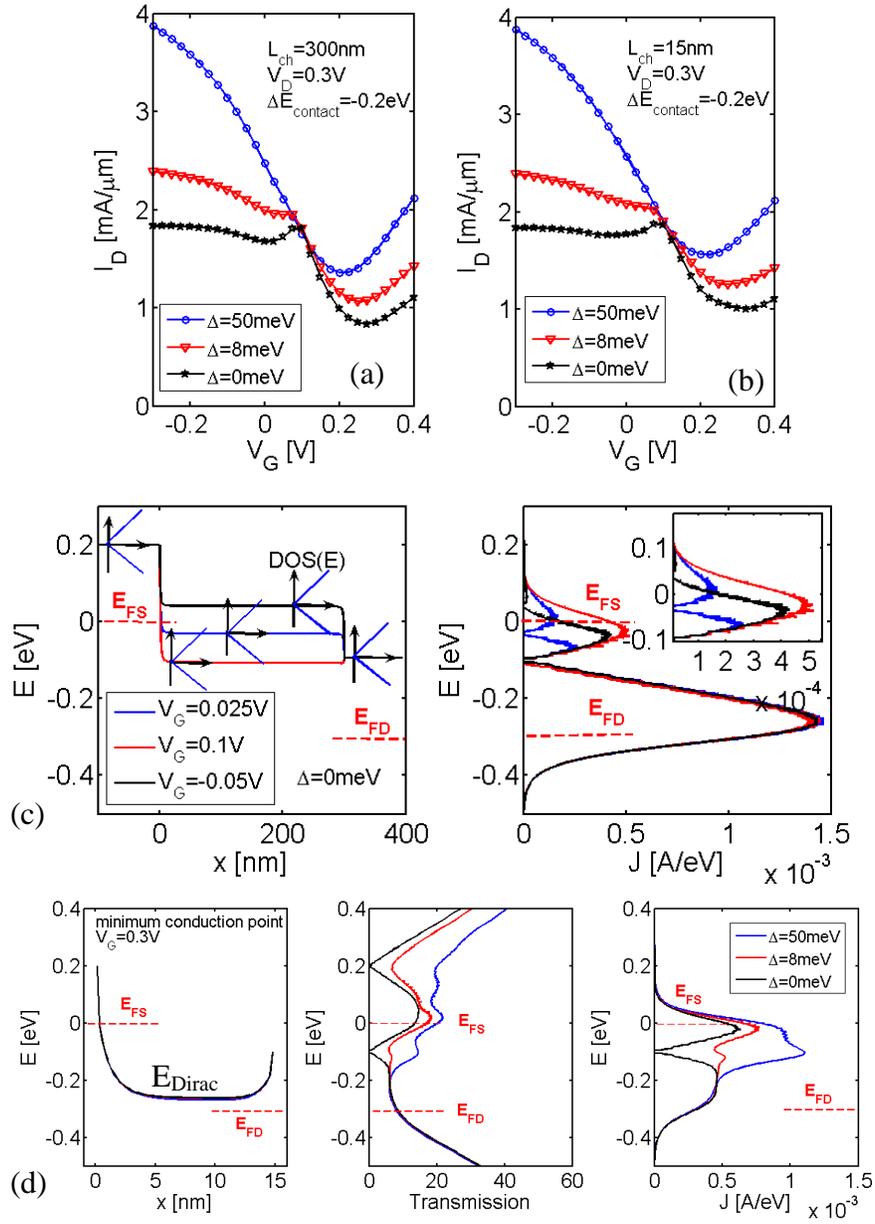

Figure 5

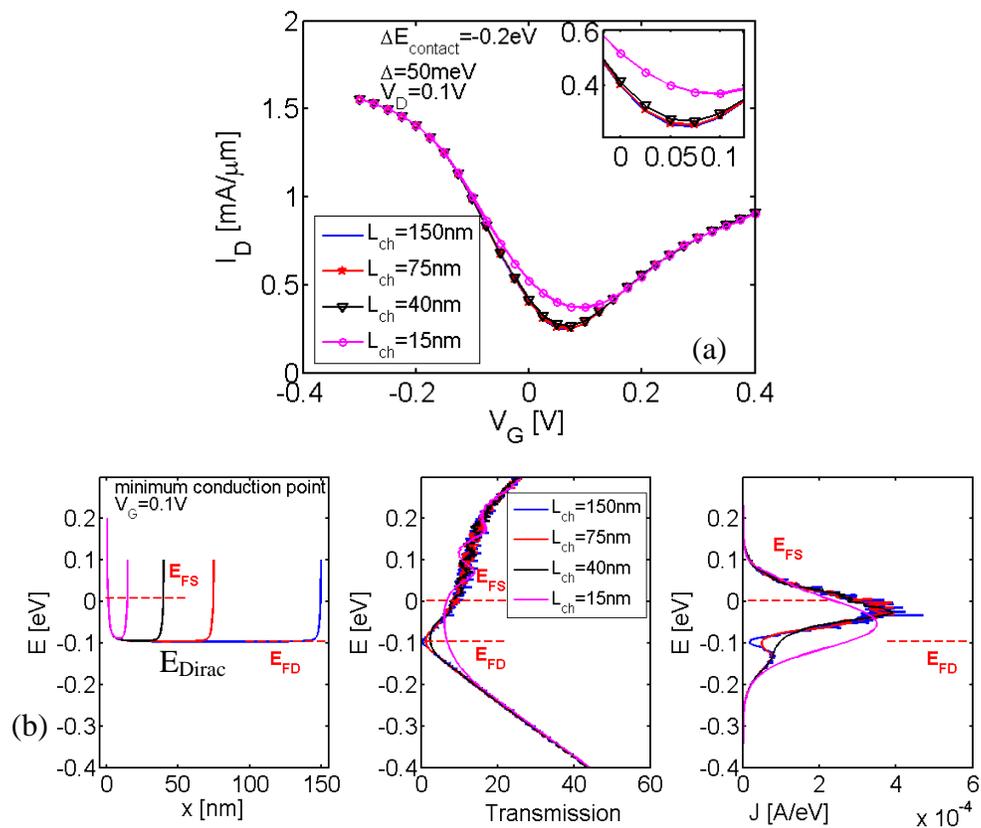

Figure 6

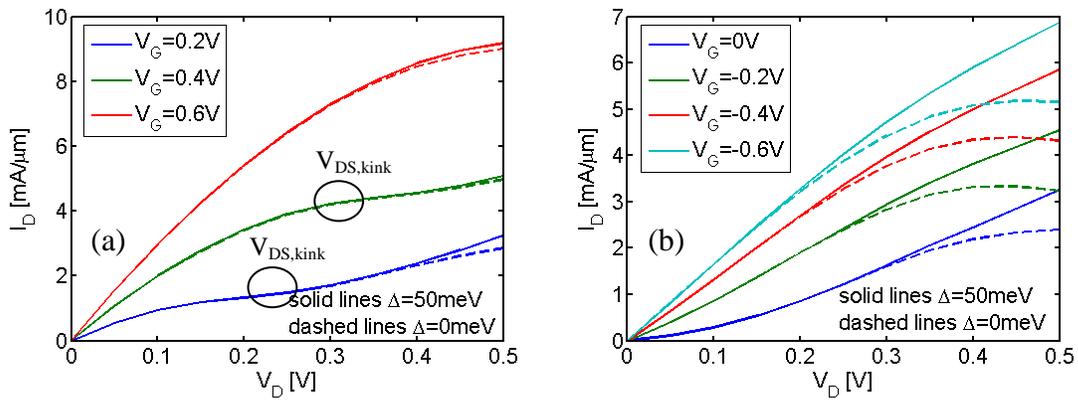

Figure 7